\documentclass[5p,times,twocolumn,authoryear]{elsarticle}

\usepackage[T1]{fontenc}
\usepackage{amsmath,amssymb}
\usepackage{graphicx}
\usepackage{natbib}
\usepackage[colorlinks=true]{hyperref}

\journal{Icarus}

\begin{document}

\begin{frontmatter}

\title{An Upper Limit on Turbulent Viscosity in the Circumjovian Nebula}

\author[caltech]{Konstantin Batygin\corref{cor1}}
\ead{kbatygin@gps.caltech.edu}
\cortext[cor1]{Corresponding author.}

\author[umich,umichastro]{Fred C. Adams}

\address[caltech]{Division of Geological and Planetary Sciences,
California Institute of Technology, Pasadena, CA 91125, USA}
\address[umich]{Department of Physics, University of Michigan,
Ann Arbor, MI 48109, USA}
\address[umichastro]{Department of Astronomy, University of Michigan,
Ann Arbor, MI 48109, USA}

\begin{abstract}
Circumplanetary disks are open, dynamic systems sustained by a meridional circulation that draws gas and dust from the parent nebula along nearly polar streamlines, and ultimately returns some of this material through a viscously driven, in-plane outflow. In such a disk, the water ice line acts as a site of solid-mass accumulation: inward-drifting icy solids sublimate interior to the ice line, the resulting vapor is advected outward by the gas, and recondensation beyond the ice line resets the material to a small Stokes number. The operation of this drift-mediated loop requires the fragmentation-limited Stokes number of the mass-dominant icy aggregates to exceed the equilibrium Stokes number at which the radial drift of solids reverses. Because the former scales inversely with the Shakura--Sunyaev viscosity parameter $\alpha$ while the latter is directly proportional to it, this requirement yields a compact upper bound on the vigor of turbulence in the satellite-forming region. Adopting an actively heated, steady-state circumjovian disk truncated at the tidal radius, we find a marginal bound of $\alpha\lesssim 10^{-3}$ for an icy-particle fragmentation threshold of $v_f\simeq1\,{\rm m\,s^{-1}}$ -- appropriate for cold ice with sticking properties similar to silicate dust. This bound scales linearly with $v_f$, reaching $\alpha\lesssim5\times10^{-3}$ for more adhesive, warm ice ($v_f\simeq5\,{\rm m\,s^{-1}}$). The bound carries no explicit dependence on the disk's mass flux, and is set entirely by the local thermal state, pressure gradient, and geometry of the circumjovian nebula.
\end{abstract}


\end{frontmatter}

\section{Introduction}

Dating back to the earliest models of Galilean satellite
formation, the accretion of the moons has been envisioned as planet
formation in miniature: solids embedded in a gas-rich disk
surrounding the growing Jupiter coagulate, settle, and assemble
into satellites
\citep{LunineStevenson1982,CanupWard2002,MosqueiraEstrada2003}.
Proposed architectures for this circumjovian nebula span a
considerable range, from the massive, quiescent minimum-mass
subnebula of \citet{MosqueiraEstrada2003} to the low-density,
gas-starved accretion disk of \citet{CanupWard2002,CanupWard2006}.
Observations of planet-driven meridional circulation and of nascent
circumplanetary disks \citep{Teague2019,Benisty2021}, together with
simulations of the gas flow that unfolds within the Hill sphere
\citep{Tanigawa2012,Szulagyi2014,Lambrechts2019,BaileyZhu2024,
Krapp2024}, have since sharpened this picture and made clear
that the circumjovian disk was not a static reservoir, but an
evolving structure -- continuously
replenished by nebular inflow and reshaped by viscous transport.
Modern treatments of satellite formation are accordingly framed
within this dynamic environment: the architecture and composition
of the satellite system reflect not only the total mass of solids
delivered to the disk, but also the thermodynamics and gas--solid
aerodynamic coupling that govern how condensable species are
transported, recycled, and concentrated within the
satellite-forming region
\citep{BatyginMorbidelli2020,AdamsBatygin2022,
AdamsBatygin2025}.

\begin{figure*}[t]
\centering
\includegraphics[width=\textwidth]{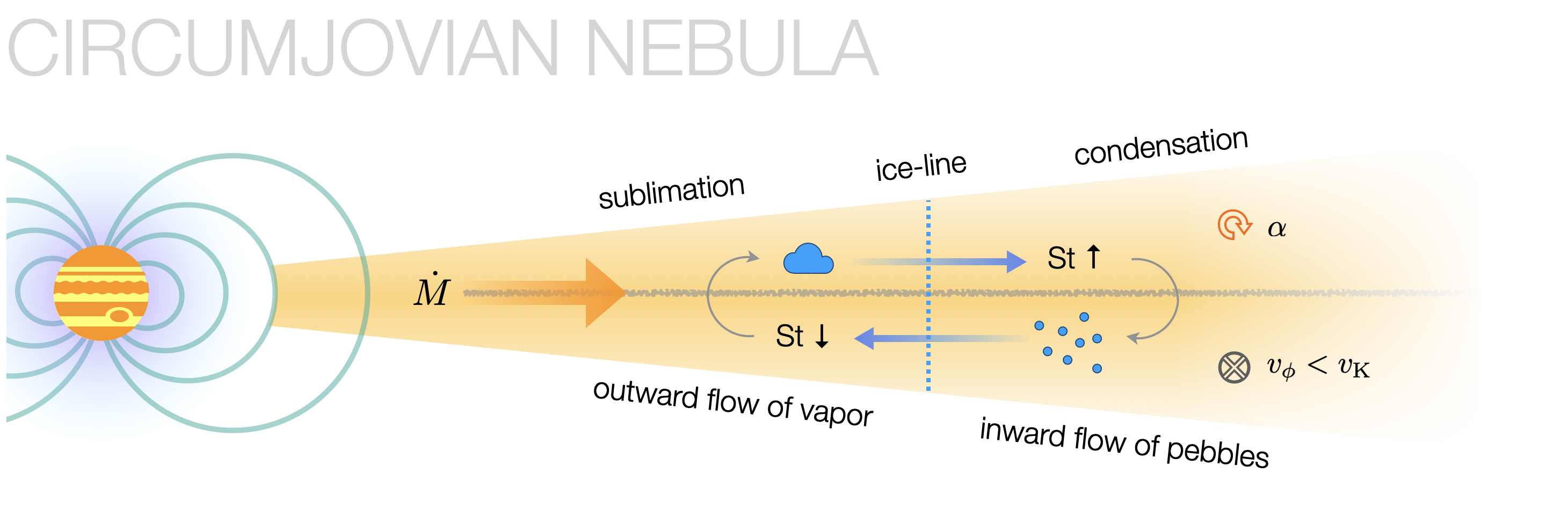}
\caption{Schematic architecture of the circumjovian nebula and the
ice-line recycling loop (not to scale). Gas is delivered to
Jupiter's Hill sphere from the parent nebula; a substantial
fraction accretes inward and is deposited onto the planet at the
magnetospheric boundary, while the remainder spreads viscously outward, carrying the
decretionary mass flux $\dot M$. Interior to the water ice line at
$r_{\rm ice}\simeq30\,R_J$ ($T_{\rm ice}\simeq170\,{\rm K}$ for
$h/r\simeq0.1$), water resides in the vapor phase and is advected
outward with the decreting gas. Upon crossing the ice line, the
vapor condenses onto small, tightly coupled grains, which coagulate to progressively larger Stokes numbers
(${\rm St}\uparrow$); once growth
carries them past the equilibrium Stokes number, the sub-Keplerian
headwind ($v_\phi<v_{\rm K}$) reverses their radial drift, and they
return to the ice line as pebbles, where sublimation
(${\rm St}\downarrow$) closes the loop. The competition between
this cycle and the intensity of the turbulence, parameterized by
$\alpha$, is quantified in Figure~\ref{fig:alpha_limit}.}
\label{fig:schematic}
\end{figure*}

A particularly important feature of this problem is the water ice
line, across which water transitions from vapor that is dynamically
tied to the gas to solid grains that can grow, drift, and settle.
In a viscously spreading circumplanetary disk
\citep{Pringle1991,BatyginMorbidelli2020}, this phase boundary
takes on a dynamical role.
Within the decretionary midplane flow of
such systems, tightly coupled particles are carried outward with
the gas, while sufficiently decoupled particles drift inward
because of the headwind associated with sub-Keplerian rotation
\citep{Weidenschilling1977,NSH1986} -- a competition that defines a
critical Stokes number at which the radial motion of solids changes
sign.
Near the ice line, sublimation and condensation continually
move material from one side of this threshold to the other: icy
particles exterior to the ice line grow by coagulation until they
decouple from the gas and drift inward; upon crossing the ice line
they sublimate; and the resulting vapor is advected back outward,
recondensing beyond the ice line as small, tightly coupled grains
whose renewed growth closes a recycling loop.
The net result is
that the ice line acts as a trap for condensable material and a
preferred site for the formation of the icy building blocks of the
Galilean satellites (Figure~\ref{fig:schematic}).

Vapor--solid exchange of this kind at the water snow line has a long
lineage in the protoplanetary-disk literature -- from the diffusive
``cold-finger'' mechanism of \citet{StevensonLunine1988}, through
models of solid enhancement at evaporation fronts
\citep{CuzziZahnle2004}, to planetesimal-forming pile-ups
in the vicinity of the snow line
\citep{IdaGuillot2016,SchoonenbergOrmel2017,DrazkowskaAlibert2017}.
The circumplanetary variant of the process differs in one essential
respect: in a decretion disk, vapor is returned across the ice line
by advection with the mean flow, rather than by turbulent diffusion
against an accretionary stream.
Recently, \citet{YapStevenson2026}
developed a detailed model of precisely this recycling process in
decreting circumplanetary disks, showing that it naturally drives
the solid population beyond the ice line toward the ice-rich
compositions exhibited by Ganymede, Callisto, and Titan.

This picture raises a simple but consequential question: what level of
turbulence is compatible with the operation of this recycling loop?
The answer is controlled by the Shakura--Sunyaev parameter $\alpha$
\citep{ShakuraSunyaev1973}.
Turbulence plays two opposing roles.
First,
it sets the viscous transport of gas and therefore the outward radial
velocity of vapor and tightly coupled particles.
Second, it controls
the turbulent velocity dispersion of solids
\citep{OrmelCuzzi2007}, and therefore the fragmentation barrier that
limits the maximum particle size \citep{Birnstiel2012}.
A larger value
of $\alpha$ strengthens the outward gas flow that particles must
overcome, but also increases collision speeds and suppresses growth to large Stokes numbers.
Requiring that particles grow large enough to
overcome outward decretion and drift back toward the ice line
therefore yields a direct upper bound on $\alpha$.

In this paper, we derive this bound for the circumjovian nebula.
The
argument is deliberately minimal.
It requires no model for the
nonlinear conversion of solids into satellitesimals, for the
subsequent dynamical assembly of the Galilean satellites, or for the
compositional evolution of the solid reservoir that recycling
produces \citep[see][]{YapStevenson2026}.
Instead, it rests on a compact criterion for the
drift-mediated recycling loop to operate in bulk: the
fragmentation-limited Stokes number of the mass-dominant icy
particles must exceed the equilibrium
Stokes number at which radial drift reverses.
Because both quantities
depend on $\alpha$ in opposite ways, this requirement translates into
a compact constraint on the turbulence level in the satellite-forming
disk.

\section{An Upper Limit on $\alpha$ in the Circumjovian Nebula}

We adopt the circumplanetary disk model of \citet{BatyginMorbidelli2020}
\citep[for analytic treatments of this family of disk models,
see][]{AdamsBatygin2022,AdamsBatygin2025},
in which gas delivered to Jupiter's Hill sphere settles into a
viscously spreading disk.
In the satellite-forming region, the
relevant midplane flow is directed outward.
This distinction is
central to the mechanism considered here: vapor and tightly coupled
condensates are carried away from the planet by the gas, while
sufficiently large particles drift inward under aerodynamic drag.

One might reasonably ask why the disk should be decreting at all,
given that Jupiter must, on the whole, gain mass.
The resolution lies in the diffusive
character of viscous evolution: material interior to a stagnation
radius ${\cal R}$ flows inward onto the planet, while material
exterior to ${\cal R}$ spreads outward.
Within the analytic
framework of \citet{AdamsBatygin2025}, this radius is
${\cal R}=u\,R_{\rm C}=u\,\lambda^2 R_{\rm H}/3$, where
$\lambda$ quantifies the angular-momentum bias of the material
delivered to the Hill sphere, $R_{\rm C}$ is the centrifugal
radius, and the dimensionless coefficient $u\simeq0.35$--$0.83$ encodes
the inflow geometry.
This budget delineates the domain of validity
of our treatment.
For
$\lambda\leqslant(3\,r_{\rm ice}/R_{\rm H})^{1/2}\simeq0.35$, the
entire source region lies interior to the water ice line
($r_{\rm ice}\simeq30\,R_J$; derived below), and the source-free
decretion solution adopted in this work provides an adequate
approximation throughout the region of interest.
For
$0.35\lesssim\lambda\lesssim[3\,r_{\rm ice}/(uR_{\rm H})]^{1/2}
\simeq0.4$--$0.6$, continued mass deposition exterior to the ice
line diminishes the local outward flux, introducing an order-unity
correction to the flow velocity.
Larger values of $\lambda$,
however, are incompatible with the picture developed here: the stagnation point then migrates beyond the ice line, and the
inward flow interior to ${\cal R}$ would sweep nascent solids into
the planet rather than allowing them to accumulate.

In the steady decretion approximation \citep{Pringle1991,Lee1991},
with the source term confined to negligibly small orbital
radii, the gas surface density can be written as
\begin{equation}
    \Sigma =
    \frac{\dot{M}}{3\pi\nu}
    \left[
    \left(\frac{\chi\,R_{\rm H}}{r}\right)^{1/2}-1
    \right],
    \label{eq:sigma_decretion}
\end{equation}
where $\dot{M}>0$ denotes the outward mass flux through the
circumplanetary disk, $\nu$ is the turbulent viscosity, and the
outer edge of the solution is placed at the tidal truncation
radius, $R_{\rm out}=\chi\,R_{\rm H}$, with $R_{\rm H}$ Jupiter's
Hill radius and $\chi\simeq0.4$ \citep{MartinLubow2011}.
Because
the solar torque extracts the disk's angular-momentum flux at this
edge, the truncated boundary plays the role of a zero-couple radius
of the classical decretion solution.
This choice departs from the
$R_{\rm H}$ boundary adopted by \citet{BatyginMorbidelli2020}.
For
convenience, we define
\begin{equation}
    {\cal B}(r)
    \equiv
    \left(\frac{\chi\,R_{\rm H}}{r}\right)^{1/2}-1 ,
    \label{eq:B_def}
\end{equation}
and note that the bound derived below scales only as
${\cal B}^{1/2}$.

Using the Shakura--Sunyaev prescription \citep{ShakuraSunyaev1973}
\begin{equation}
    \nu = \alpha \frac{c_s^2}{\Omega},
    \label{eq:ss_alpha}
\end{equation}
the radial velocity of the gas follows from mass conservation:
\begin{equation}
    v_{r,{\rm gas}}
    =
    \frac{\dot{M}}{2\pi r\Sigma}
    =
    \frac{3\nu}{2r{\cal B}}
    =
    \frac{3}{2}
    \frac{\alpha}{{\cal B}}
    \left(\frac{c_s}{v_K}\right)^2
    v_K ,
    \label{eq:vr_gas}
\end{equation}
where $v_K=r\Omega$ is the Keplerian velocity.
The sign of
equation~(\ref{eq:vr_gas}) is positive, corresponding to outward
mean flow.
Throughout, we adopt the standard single-$\alpha$
closure, in which the same parameter characterizes
angular-momentum transport, turbulent particle stirring, and
particle diffusion (i.e., the turbulent Schmidt number is taken to
be unity), as is customary in calculations of the early
stages of dust evolution
\citep{Birnstiel2012,DrazkowskaSzulagyi2018}.\footnote{If the
stress and collisional-stirring parameters are permitted to differ,
the bound derived below applies to their geometric mean,
$(\alpha_\nu\,\alpha_{\rm coll})^{1/2}$.}

The radial velocity of a solid particle with Stokes number ${\rm St}$
is \citep{Weidenschilling1977,NSH1986}
\begin{equation}
    v_{r,{\rm solid}}
    =
    \frac{
    v_{r,{\rm gas}}
    -
    2\eta v_K {\rm St}
    }{
    1+{\rm St}^2
    },
    \label{eq:vr_solid}
\end{equation}
where $\eta v_K$ is the sub-Keplerian headwind speed.
Writing
\begin{equation}
    \eta = \xi\left(\frac{c_s}{v_K}\right)^2,
    \label{eq:eta_def}
\end{equation}
the headwind coefficient is set by the local radial pressure
gradient,
\begin{equation}
    \xi = -\frac{1}{2}\,
    \frac{\partial \ln P}{\partial \ln r},
    \label{eq:xi_def}
\end{equation}
evaluated at the disk midplane.
In the active, optically thin
steady state of the \citet{BatyginMorbidelli2020} model, the aspect
ratio is approximately constant at $h/r\simeq0.1$ (so that
$T\propto1/r$)
and the surface density is well described by
$\Sigma\propto r^{-5/4}$.
The midplane pressure then scales as
$P\propto\Sigma\,c_s\,\Omega\propto r^{-13/4}$, and the headwind
coefficient takes the value\footnote{The adopted
$\Sigma\propto r^{-5/4}$ is a global power-law fit to the decretion
solution of equation~(\ref{eq:sigma_decretion}). With
$\nu\propto r^{1/2}$ (as follows from constant $h/r$), the local
logarithmic slope of that solution at the ice line is
${\rm d}\ln\Sigma/{\rm d}\ln r=-1/2-s/[2(s-1)]\simeq-1.23$, where
$s\equiv(\chi R_{\rm H}/r_{\rm ice})^{1/2}\simeq3.14$; the
corresponding headwind coefficient is $\xi=1.62$, a difference that
propagates to the final bound at the sub-percent level.}
\begin{equation}
    \xi=\frac{13}{8}.
    \label{eq:xi_BM}
\end{equation}
Equation~(\ref{eq:vr_solid}) implies the existence of an equilibrium
Stokes number at which the radial motion of solids vanishes.
Setting
$v_{r,{\rm solid}}=0$ gives
\begin{equation}
    {\rm St}_{\rm eq}
    =
    \frac{v_{r,{\rm gas}}}{2\eta v_K}
    =
    \frac{3}{4}
    \frac{\alpha}{\xi{\cal B}} .
    \label{eq:St_eq}
\end{equation}
Equation~(\ref{eq:St_eq}) recovers the sign dichotomy described in
Section 1: particles below ${\rm St}_{\rm eq}$ ride the
decretionary outflow, while those above it drift inward -- the
exchange across this threshold constituting the recycling loop.

For this cycle to operate, particles exterior to the ice line must
grow to at least ${\rm St}_{\rm eq}$ -- a requirement that
constrains the turbulent state of the disk.
In a turbulent gas,
the relative velocity of small particles scales as
\citep{OrmelCuzzi2007}
\begin{equation}
    \Delta v_{\rm turb}
    \sim
    \sqrt{\alpha\, {\rm St}}\,c_s ,
    \label{eq:turb_vel}
\end{equation}
up to order-unity coefficients.\footnote{Equation~(\ref{eq:turb_vel})
holds in the intermediate regime of \citet{OrmelCuzzi2007}, wherein
the particle stopping time exceeds the turnover time of the
smallest eddies: ${\rm St}>{\rm Re}^{-1/2}$, with
${\rm Re}=\nu/\nu_{\rm mol}$ the turbulent Reynolds number and
$\nu_{\rm mol}$ the molecular viscosity. Because
$\Sigma\propto\dot M/\nu$ in the steady decretion solution,
$\alpha$ cancels between the two viscosities, leaving
${\rm Re}\simeq2\times10^{8}\,(\dot M/0.1\,M_J\,{\rm Myr}^{-1})$
at the ice line. The condition is most stringent at the marginal
point, where ${\rm St}_{\rm eq}={\rm St}_{\max}=
(v_f/2c_s)\sqrt{\zeta/(\xi{\cal B})}\simeq1.1\times10^{-3}$ and
$2.1\times10^{-4}$ for $v_f=5$ and $1\,{\rm m\,s^{-1}}$, so that
${\rm Re}>{\rm St}^{-2}$ translates to
$\dot M\gtrsim4\times10^{-4}$ and
$\dot M\gtrsim10^{-2}\,M_J\,{\rm Myr}^{-1}$, respectively. This
requirement -- the sole avenue by which $\dot M$ enters the bound
of equation~(\ref{eq:alpha_bound_general}) -- is comfortably
satisfied by the fiducial decretionary mass flux of
$\dot M\simeq0.1\,M_J\,{\rm Myr}^{-1}$ adopted by
\citet{BatyginMorbidelli2020} (see also
\citealp{BatyginAdams2025}).} Growth
stalls once turbulent collision
speeds approach the fragmentation threshold $v_f$.
We therefore write
the fragmentation-limited Stokes number of the mass-dominant
aggregates as \citep{Birnstiel2012}
\begin{equation}
    {\rm St}_{\max}
    =
    \frac{\zeta}{3}
    \frac{v_f^2}{\alpha c_s^2},
    \label{eq:St_max}
\end{equation}
where $\zeta\lesssim1$ absorbs order-unity uncertainties associated
with the turbulent collision model and the particle-size
distribution; the numerical value $\zeta=0.37$ adopted here
coincides with the fragmentation-barrier calibration factor of
\citet{Birnstiel2012}.
This calibration tracks the mass-dominant,
flux-carrying size of the fragmentation-limited distribution -- the
quantity of relevance for a mass-recycling loop -- rather than the
absolute upper boundary of the size distribution.
We note, however,
that adopting the latter ($\zeta\rightarrow1$) loosens the bound
derived below by only a factor of $(0.37)^{-1/2}\simeq1.6$.

The operational criterion for drift-mediated recycling of the bulk
condensable mass is
\begin{equation}
    {\rm St}_{\max}>{\rm St}_{\rm eq}.
    \label{eq:condition}
\end{equation}
Substituting equations~(\ref{eq:St_eq}) and~(\ref{eq:St_max}) yields
\begin{equation}
    \frac{\zeta}{3}
    \frac{v_f^2}{\alpha c_s^2}
    >
    \frac{3}{4}
    \frac{\alpha}{\xi{\cal B}},
    \label{eq:ineq1}
\end{equation}
or
\begin{equation}
    \alpha^2
    <
    \frac{4}{9}
    \zeta \xi {\cal B}
    \frac{v_f^2}{c_s^2}.
    \label{eq:ineq2}
\end{equation}
Thus, the turbulence level must satisfy
\begin{equation}
    \boxed{
    \alpha
    <
    \frac{2}{3}
    \frac{v_f}{c_s}
    \left(\zeta \xi {\cal B}\right)^{1/2}
    }
    \label{eq:alpha_bound_general}
\end{equation}
for icy particles to defeat outward advection and close the
drift-mediated sublimation--condensation recycling loop.
With
$\xi=13/8$ from equation~(\ref{eq:xi_BM}), the bound reduces to
\begin{equation}
    \alpha
    <
    \frac{2}{3}\sqrt{\frac{13\,\zeta\,{\cal B}}{8}}\,\frac{v_f}{c_s}.
    \label{eq:alpha_bound_BM}
\end{equation}
Equation~(\ref{eq:alpha_bound_BM}) is a {\it marginal}
condition: at saturation, the mass-dominant fragmentation-limited
particles have ${\rm St}_{\max}={\rm St}_{\rm eq}$ and vanishing radial
velocity, so net inward transport requires $\alpha$ to lie below this bound
with some margin (Section 3).

We note that equation~(\ref{eq:alpha_bound_general}) carries no
explicit dependence on the absolute mass flux $\dot M$ through the
circumplanetary disk: because both the surface density and the
radial gas velocity are evaluated within the same steady decretion
solution, $\dot M$ cancels identically at fixed thermal structure and
ice-line location, leaving the bound set by the local thermal
state, pressure gradient, and disk geometry.

The geometric factor ${\cal B}$ can be evaluated at the water ice
line.
The disk aspect ratio is
\begin{equation}
    \frac{h}{r}
    =
    \frac{c_s}{v_K}
    =
    \left(
    \frac{c_s^2 r}{GM_J}
    \right)^{1/2},
    \label{eq:aspect_ratio}
\end{equation}
so the radius corresponding to a given ice-line sound speed is
\begin{equation}
    r_{\rm ice}
    =
    \left(\frac{h}{r}\right)^2
    \frac{GM_J}{c_s^2}.
    \label{eq:rice}
\end{equation}
At the water ice line, $T_{\rm ice}\simeq 170\,{\rm K}$,
giving $c_s=(k_BT_{\rm ice}/\mu)^{1/2}\simeq0.77\,{\rm
km\,s^{-1}}$ for $\mu\simeq2.34\,m_p$.
For $h/r\simeq0.1$ -- a value
independently required by the dynamics of Amalthea's inward
resonant transport by Io, which demands $h/r\gtrsim0.08$ during
the epoch of satellite formation \citep{BruntonBatygin} --
equation~(\ref{eq:rice}) gives $r_{\rm ice}\simeq 30\,R_J$, just
beyond the present orbit of Callisto\footnote{Intriguingly,
the dynamical reconstruction of the Galilean satellites' resonant
architecture by \citet{YapBatygin2026} invokes a pressure maximum
in the vicinity of Callisto's present-day orbit, which functions
as a migration trap that excludes Callisto from the Laplace
resonance; the coincidence of this feature with the water ice line
derived here is suggestive.} ($a\simeq26\,R_J$).
The residence of ice-rich Ganymede and Callisto interior to this
radius is naturally accommodated: satellitesimals form at or beyond
the ice line and are subsequently drawn inward by gas-assisted
migration
\citep{BatyginMorbidelli2020,BruntonBatygin}.
Adopting $R_{\rm H}\simeq740\,R_J$ for Jupiter and $\chi\simeq0.4$,
the corresponding value of the geometric factor is
${\cal B}_{\rm ice}=(\chi R_{\rm H}/r_{\rm ice})^{1/2}-1\simeq2.1$.

\begin{figure}[!ht]
\centering
\includegraphics[width=\columnwidth]{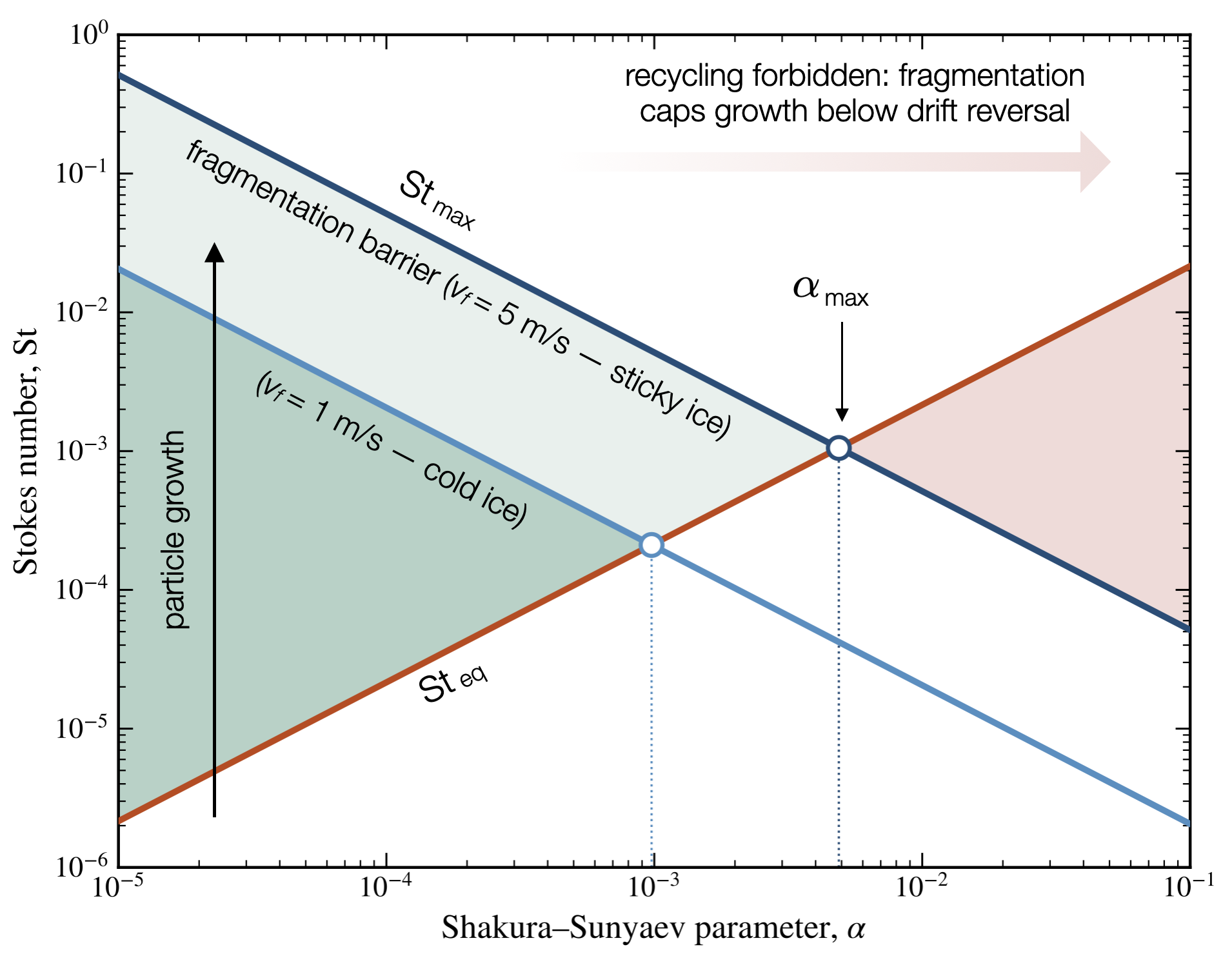}
\caption{The turbulence constraint at the circumjovian ice line, in
the $(\alpha,\,{\rm St})$ plane. The mass-dominant,
fragmentation-limited Stokes number
${\rm St}_{\max}=(\zeta/3)\,v_f^2/(\alpha c_s^2)$ is shown for both
fragmentation branches (dark blue: sticky ice,
$v_f=5\,{\rm m\,s^{-1}}$; light blue: cold ice,
$1\,{\rm m\,s^{-1}}$), falling with increasing turbulence, while
the drift-reversal threshold
${\rm St}_{\rm eq}=(3/4)\,\alpha/(\xi{\cal B})$ (red) rises. In the
dark green wedge, drift-mediated recycling operates for either
branch; in the light green wedge, it additionally requires sticky
ice and sufficiently dust-laden inflow
($Z_{\rm in}\gtrsim2.5\times10^{-3}$ for the full window;
equation~\ref{eq:locality}).
The windows are pinched off at
$\alpha_{\max}=(2/3)(v_f/c_s)(\zeta\xi{\cal B})^{1/2}
\simeq5\times10^{-3}$ and $10^{-3}$ (open circles),
beyond which fragmentation caps the growth of the mass-dominant
population below the drift-reversal threshold and bulk
drift-mediated recycling is suppressed (red wedge).
Note that systematic drift dominates over turbulent diffusion
of solids only for $\alpha\lesssim0.25\,\alpha_{\max}$
(equation~\ref{eq:drift_vs_diff}), so the efficiently operating
regime lies well inside each wedge.
All quantities are evaluated at the ice line
($T_{\rm ice}\simeq170\,{\rm K}$, $r_{\rm ice}\simeq30\,R_J$,
${\cal B}_{\rm ice}\simeq2.1$); see Figure~\ref{fig:schematic} for
the physical setup.}
\label{fig:alpha_limit}
\end{figure}

The remaining input is the fragmentation threshold of icy
aggregates.
Laboratory and numerical studies span values of order
$1$--$10\,{\rm m\,s^{-1}}$, depending on temperature, porosity,
monomer size, and aggregate structure
\citep{Wada2009,GundlachBlum2015,MusiolikWurm2019,Gundlach2018}.
In
practice, recondensation beyond the ice line proceeds
heterogeneously, onto the pre-existing population of rocky grains;
because the condensable mass is comparable to or exceeds the
refractory mass, the fragmentation threshold remains that of
ice-mantled aggregates, although silicate cores may render such grains marginally more
robust, loosening the bound only modestly.
With these inputs,
equation~(\ref{eq:alpha_bound_BM}) evaluates to
\begin{equation}
    \alpha
    \lesssim
    10^{-3}
    \left(\frac{v_f}{1\,{\rm m\,s^{-1}}}\right),
    \label{eq:alpha_bound_num}
\end{equation}
such that the $v_f\simeq1\,{\rm m\,s^{-1}}$ cold-ice value
defines the operational limit, while
sticky ice ($v_f\simeq5\,{\rm m\,s^{-1}}$) loosens the bound to
$\alpha\lesssim5\times10^{-3}$.
It is further worth noting that multiwavelength modeling of
protoplanetary-disk continuum emission independently favors
comparably fragile aggregates, with
thresholds $v_f\lesssim1\,{\rm m\,s^{-1}}$ and values as low as
$0.3\,{\rm m\,s^{-1}}$ admissible \citep{Ueda2024}. The latter
would tighten the operational limit to
$\alpha\lesssim3\times10^{-4}$.
The geometry of this constraint
is summarized in Figure~\ref{fig:alpha_limit}.
If $\alpha$ exceeds
this bound, turbulent fragmentation prevents the mass-dominant icy
particles from reaching the equilibrium Stokes number required for
inward drift.
Condensates then remain tightly coupled to the outward decretionary
flow, and are eventually fed back into the parent circumstellar
disk.

The choice between the two branches of
equation~(\ref{eq:alpha_bound_num}) warrants scrutiny.
Laboratory
measurements indicate that the surface energy of water ice drops
sharply below a sticking transition at $T\simeq175$--$200\,{\rm K}$,
rendering colder ice no stickier than silicate dust
\citep{MusiolikWurm2019,Gundlach2018}; because the ice line itself
sits at $T_{\rm ice}\simeq170\,{\rm K}$ -- and the mapping from surface energy to fragmentation velocity
carries additional dependence on monomer size, porosity, and
collision geometry -- the cold value may apply even to aggregates that
coagulate in place.
Transport considerations sharpen this point:
freshly recondensed grains are carried outward, toward lower
temperatures, while they grow.
The coagulation timescale is
$t_{\rm grow}\simeq N/(Z_{\rm in}\Omega)$, where $Z_{\rm in}$ is
the solid abundance of the gas delivered to the disk and
$N=\ln(a_{\rm eq}/a_0)\simeq7$ is the number of growth $e$-folds
separating micron-sized seeds ($a_0\simeq1\,\mu$m) from the
drift-reversal size
$a_{\rm eq}=\dot M\Omega/(2\pi^2\xi\rho_{\rm s}c_s^2)\simeq1$~mm
(evaluated\footnote{The result is independent of both $\alpha$ and
${\cal B}$; at higher mass flux, the transition to Stokes drag
alters this size estimate without affecting the Stokes-number
criterion itself.} at $\dot M=0.1\,M_J\,{\rm Myr}^{-1}$ for a
material density $\rho_{\rm s}\simeq1\,{\rm g\,cm^{-3}}$).
The
fractional distance covered before drift reversal is then
\begin{equation}
    \frac{\Delta r}{r}
    \simeq
    \frac{v_{r,{\rm gas}}\,t_{\rm grow}}{r}
    =
    \frac{3}{2}
    \frac{\alpha}{{\cal B}}
    \left(\frac{h}{r}\right)^{2}
    \frac{N}{Z_{\rm in}}.
    \label{eq:locality}
\end{equation}

Requiring $\Delta r/r\lesssim0.1$ (i.e., $T\gtrsim155\,{\rm K}$)
shows that growth remains confined to the warm annulus adjacent to
the ice line only for $\alpha\lesssim2\,Z_{\rm in}$.
The two limits
of equation~(\ref{eq:alpha_bound_num}) therefore apply in the
corresponding regimes of delivered metallicity.
For dust-laden
inflow, $Z_{\rm in}\gtrsim2.5\times10^{-3}$, growth is local
across the entire warm window, and the sticky-ice limit
$\alpha\lesssim5\times10^{-3}$ applies -- provided, per the
laboratory caveat above, that ice at $T_{\rm ice}$ retains its
enhanced adhesion.
For strongly depleted inflow,
$Z_{\rm in}\lesssim5\times10^{-4}$ -- the expected circumstance at
late times, when the solids of the parent nebula have largely
settled into a midplane pebble layer inaccessible to the polar
streamlines that feed the circumplanetary disk -- any $\alpha$ in
excess of the cold limit carries growing grains beyond the warm
annulus, so the sticky-ice branch cannot be invoked, and
$\alpha\lesssim10^{-3}$ holds irrespective of the
sticking properties of warm ice.
Intermediate metallicities bridge
the two regimes, with $\alpha_{\max}\simeq2\,Z_{\rm in}$; because
$N$ depends on $\dot M$ only logarithmically, these thresholds
shift by only tens of percent across the plausible range of mass
flux.
This division is, moreover, conservative with respect to the cold
branch: the laboratory
sticking transition lies at or above $T_{\rm ice}$ itself, and
outward advection during growth can only cool the environment
further, so the cold limit may well encroach on the nominally warm
regime.
The warm-ice branch is accordingly best regarded as a
permissive envelope, with the operational limit for the
anticipated dust-depleted inflow being $\alpha\lesssim10^{-3}$.

\section{Discussion and Conclusions}

The requirement that fragmentation-limited icy aggregates outgrow
the equilibrium Stokes number at which outward decretion balances
inward drift yields a closed-form upper limit on circumjovian
turbulence, applicable wherever the inflow's angular-momentum
budget renders the ice-line region decretionary (Section 2):
$\alpha\lesssim5\times10^{-3}$ for sticky ice, tightening to
$\alpha\lesssim10^{-3}$ when, under dust-depleted inflow,
grains are carried beyond the warm-ice sticking transition before
reaching the drift-reversal size (equation~\ref{eq:locality}).
Three qualifications frame this
result.
First, the condition is necessary rather than sufficient:
at saturation the mass-dominant particles have vanishing radial
velocity, so efficient recycling requires $\alpha$ substantially
below the marginal value.
Second, the bound does not by itself guarantee that freshly
recondensed grains regrow to ${\rm St}_{\rm eq}$ before being
advected away. The outward displacement accrued during regrowth is,
however, quantified by equation~(\ref{eq:locality}): growth to the
drift-reversal size completes within a small fraction of
$r_{\rm ice}$ whenever $\alpha\lesssim2\,Z_{\rm in}$, and
heterogeneous condensation onto pre-existing grains affords a
considerable head start. A full treatment nonetheless requires
coupled coagulation--transport modeling
\citep{Birnstiel2012,DrazkowskaSzulagyi2018,Shibaike2017}.
Third,
the bound is an onset condition, evaluated in the trace-particle
limit: once the trap matures and the solid loading approaches
unity, back-reaction weakens the headwind (while modifying the
gas's own radial flow) and raises the effective drift-reversal Stokes number, so the mature
loop self-limits \citep[see also][]{YapStevenson2026}.

Reducing turbulence to a single mean-field $\alpha$ likewise
excludes, by construction, the self-organization observed in
simulations with self-consistent turbulence and dust--gas feedback,
in which zonal flows, vortices, and the streaming instability
concentrate solids into dense clumps whose internal collision
velocities decouple from the global transport coefficient
\citep{Johansen2007,Johansen2011,Yang2018,XuBai2022,HuangBai2025,
TominagaTanaka2025}. Concentration of this kind could, in
principle, deliver decoupled solids to the ice line even where the
mean-field bound is violated. Such mechanisms, however, generally
require moderately coupled seeds: strong clumping becomes
difficult below ${\rm St}\sim10^{-2}$ unless the solid loading
substantially exceeds that anticipated for late-stage inflow
\citep{LiYoudin2021}. Assembling such seeds from the tightly
coupled, freshly recondensed population is precisely the
fragmentation-limited growth problem constrained here.

Systematic uncertainties enter our calculations
multiplicatively: identifying the fragmentation cap with the
absolute upper boundary of the size distribution
($\zeta\rightarrow1$) rather than its mass-dominant calibration
loosens the bounds by a factor of $1.6$, and the vertical
structure of the viscous radial flow \citep{TakeuchiLin2002}
introduces an additional order-unity correction.
Here we adopt the late-stage, optically thin disk model. If
solid enrichment instead renders the ice-line region optically
thick, its location must be recomputed from the balance of viscous
heating and radiative cooling. Such a state, if reached, would
belong to the later, feedback-dominated evolution of the trap
rather than to the onset addressed by our criterion. Moderate
displacements of the ice line affect the bound only through
${\cal B}^{1/2}$, so the numerical result remains robust at the
level of a factor of a few.
The constraint is therefore local: it neither precludes
different levels of turbulence elsewhere in the system or at other
epochs, nor bears on the thermally ionized, magnetically active
region that may operate close to the planet
\citep{Fujii2014,TurnerLeeSano2014}.

In addition to the advective transport considered here,
turbulent diffusion provides a residual exchange channel across
the ice line.
Condensation builds a concentration
peak just exterior to $r_{\rm ice}$, so the down-gradient diffusive
flux across its inner edge is directed inward at any $\alpha$; over
a front of width ${\sim}h$, the P\'eclet number
${\rm Pe}_h=\tfrac{3}{2}\,(h/r)/{\cal B}\simeq0.07$ is small, and
this diffusive leg operates even when fragmentation caps growth
below ${\rm St}_{\rm eq}$ \citep[see also][]{YapStevenson2026}.
The relevant comparison, however, is with the systematic
drift governed by condition~(\ref{eq:condition}).
Since the outward gas
advection can be written $v_{r,{\rm gas}}=2\eta v_K\,{\rm St}_{\rm
eq}$ (equation~\ref{eq:St_eq}), the net inward speed of the
mass-dominant aggregates is
$|v_{r,{\rm solid}}|=2\eta v_K({\rm St}_{\max}-{\rm St}_{\rm eq})$;
its ratio to the diffusive speed scale $\nu/h$ is
\begin{equation}
    \frac{|v_{r,{\rm solid}}|}{\nu/h}
    =
    \frac{3}{2}\,
    \frac{(h/r)}{{\cal B}}
    \left[\left(\frac{\alpha_{\max}}{\alpha}\right)^{2}-1\right],
    \label{eq:drift_vs_diff}
\end{equation}
which vanishes at the marginal boundary, exceeds unity for
$\alpha\lesssim0.25\,\alpha_{\max}$, and at $\alpha\sim10^{-4}$
reaches ${\sim}10$ (cold ice) to ${\sim}10^{2}$ (sticky ice).
Drift
therefore dominates throughout the efficiently operating regime and
is extinguished smoothly as $\alpha\rightarrow\alpha_{\max}$,
beyond which only diffusive exchange survives.

The resulting limit lends analytic support to several independent
lines of evidence for a quiescent circumjovian disk.
Magnetohydrodynamic considerations indicate that the
satellite-forming region was magnetically dead, with the
magnetorotational instability suppressed by low ionization
\citep{Fujii2014,TurnerLeeSano2014}; these studies do not fix
$\alpha$, but the absence of magnetized turbulence removes the most
efficient known driver of transport, and in such dead zones purely
hydrodynamic mechanisms -- most notably the vertical shear
instability -- supply only gentle residual stirring,
$\alpha\sim10^{-5}$--$10^{-4}$ near the midplane
\citep{NelsonGresselUmurhan2013,StollKley2014}.
Numerical models of
dust evolution in circumplanetary disks likewise find efficient
trapping of solids, albeit at the centrifugal radius, where
inflowing and outflowing gas meet \citep{DrazkowskaSzulagyi2018}.
Independent pebble-delivery and growth calculations likewise
favor weak turbulence \citep{Ronnet2017,Shibaike2017}, with
Galilean-satellite pebble-accretion models typically adopting
$\alpha\sim10^{-4}$ \citep{Shibaike2019,BatyginMorbidelli2020}.
The compositional consequences of this
recycling have been quantified by \citet{YapStevenson2026}; the
present analysis supplies the complementary statement of when
drift-mediated transport is permitted.
Within this framework,
convergent ice-line concentration in decreting circumplanetary
disks remains dynamically accessible, provided the turbulence lies
safely below the bound established here.

\section*{Acknowledgments}

We thank the two anonymous referees for their constructive
reviews that helped improve the manuscript.
We are grateful to Tony Yap, Ian Brunton, Luke Handley, and
Alessandro Morbidelli for insightful discussions. KB is grateful to
the David and Lucile Packard Foundation, the Caltech Center for
Comparative Planetary Evolution (3CPE), the National Science
Foundation (grant number: AST 2408867) and NASA (Emerging Worlds
grant number: 80NSSC26K0395) for their generous support. FCA
acknowledges support from NSF Grant No.~2508843.

\section*{Declaration of competing interest}

The authors declare that they have no known competing financial
interests or personal relationships that could have appeared to
influence the work reported in this paper.

\section*{Data availability}

No new data were generated or analyzed in support of this research.



\begin{thebibliography}{99}

\bibitem[Adams and Batygin(2022)]{AdamsBatygin2022}
Adams, F.~C., Batygin, K., 2022.
Analytic approach to the late stages of giant planet formation.
The Astrophysical Journal 934, 111.

\bibitem[Adams and Batygin(2025)]{AdamsBatygin2025}
Adams, F.~C., Batygin, K., 2025.
General analytic solutions for circumplanetary disks during the late
stages of giant planet formation.
Publications of the Astronomical Society of the Pacific 137, 054401.

\bibitem[Bailey and Zhu(2024)]{BaileyZhu2024}
Bailey, A.~P., Zhu, Z., 2024.
Growing planet envelopes in spite of recycling flows.
Monthly Notices of the Royal Astronomical Society 534, 2953--2967.

\bibitem[Batygin and Adams(2025)]{BatyginAdams2025}
Batygin, K., Adams, F.~C., 2025.
Determination of Jupiter's primordial physical state.
Nature Astronomy 9, 835--844.

\bibitem[Batygin and Morbidelli(2020)]{BatyginMorbidelli2020}
Batygin, K., Morbidelli, A., 2020.
Formation of giant planet satellites.
The Astrophysical Journal 894 (2), 143.

\bibitem[Benisty et al.(2021)]{Benisty2021}
Benisty, M., Bae, J., Facchini, S., et al., 2021.
A circumplanetary disk around PDS 70c.
The Astrophysical Journal Letters 916, L2.

\bibitem[Birnstiel et al.(2012)]{Birnstiel2012}
Birnstiel, T., Klahr, H., Ercolano, B., 2012.
A simple model for the evolution of the dust population in
protoplanetary disks.
Astronomy \& Astrophysics 539, A148.

\bibitem[Brunton and Batygin(2025)]{BruntonBatygin}
Brunton, I.~R., Batygin, K., 2025.
On the origin and dynamical evolution of Jupiter's moon Amalthea.
The Astrophysical Journal 991, 15.

\bibitem[Canup and Ward(2002)]{CanupWard2002}
Canup, R.~M., Ward, W.~R., 2002.
Formation of the Galilean satellites: conditions of accretion.
The Astronomical Journal 124, 3404--3423.

\bibitem[Canup and Ward(2006)]{CanupWard2006}
Canup, R.~M., Ward, W.~R., 2006.
A common mass scaling for satellite systems of gaseous planets.
Nature 441, 834--839.

\bibitem[Cuzzi and Zahnle(2004)]{CuzziZahnle2004}
Cuzzi, J.~N., Zahnle, K.~J., 2004.
Material enhancement in protoplanetary nebulae by particle drift
through evaporation fronts.
The Astrophysical Journal 614, 490--496.

\bibitem[Dr\k{a}\.zkowska and Alibert(2017)]{DrazkowskaAlibert2017}
Dr\k{a}\.zkowska, J., Alibert, Y., 2017.
Planetesimal formation starts at the snow line.
Astronomy \& Astrophysics 608, A92.

\bibitem[Dr\k{a}\.zkowska and Szul\'agyi(2018)]{DrazkowskaSzulagyi2018}
Dr\k{a}\.zkowska, J., Szul\'agyi, J., 2018.
Dust evolution and satellitesimal formation in circumplanetary disks.
The Astrophysical Journal 866, 142.

\bibitem[Fujii et al.(2014)]{Fujii2014}
Fujii, Y.~I., Okuzumi, S., Tanigawa, T., Inutsuka, S., 2014.
On the viability of the magnetorotational instability in
circumplanetary disks.
The Astrophysical Journal 785, 101.

\bibitem[Gundlach and Blum(2015)]{GundlachBlum2015}
Gundlach, B., Blum, J., 2015.
The stickiness of micrometer-sized water-ice particles.
The Astrophysical Journal 798, 34.

\bibitem[Gundlach et al.(2018)]{Gundlach2018}
Gundlach, B., Schmidt, K.~P., Kreuzig, C., Bischoff, D., Rezaei, F.,
Kothe, S., Blum, J., Grzesik, B., Stoll, E., 2018.
The tensile strength of ice and dust aggregates and its dependence on
particle properties.
Monthly Notices of the Royal Astronomical Society 479, 1273--1277.

\bibitem[Huang and Bai(2025)]{HuangBai2025}
Huang, P., Bai, X.-N., 2025.
Dust clumping in outer protoplanetary disks: the interplay among
four instabilities.
The Astrophysical Journal Letters 986, L13.

\bibitem[Ida and Guillot(2016)]{IdaGuillot2016}
Ida, S., Guillot, T., 2016.
Formation of dust-rich planetesimals from sublimated pebbles inside
of the snow line.
Astronomy \& Astrophysics 596, L3.

\bibitem[Johansen et al.(2007)]{Johansen2007}
Johansen, A., Oishi, J.~S., Mac Low, M.-M., Klahr, H., Henning, T.,
Youdin, A., 2007.
Rapid planetesimal formation in turbulent circumstellar disks.
Nature 448, 1022--1025.

\bibitem[Johansen et al.(2011)]{Johansen2011}
Johansen, A., Klahr, H., Henning, T., 2011.
High-resolution simulations of planetesimal formation in turbulent
protoplanetary discs.
Astronomy \& Astrophysics 529, A62.

\bibitem[Krapp et al.(2024)]{Krapp2024}
Krapp, L., Kratter, K.~M., Youdin, A.~N., Ben\'itez-Llambay, P.,
Masset, F., Armitage, P.~J., 2024.
A thermodynamic criterion for the formation of circumplanetary
disks.
The Astrophysical Journal 973, 153.

\bibitem[Lambrechts et al.(2019)]{Lambrechts2019}
Lambrechts, M., Lega, E., Nelson, R.~P., Crida, A., Morbidelli, A.,
2019.
Quasi-static contraction during runaway gas accretion onto giant
planets.
Astronomy \& Astrophysics 630, A82.

\bibitem[Lee et al.(1991)]{Lee1991}
Lee, U., Saio, H., Osaki, Y., 1991.
Viscous excretion discs around Be stars.
Monthly Notices of the Royal Astronomical Society 250, 432--437.

\bibitem[Li and Youdin(2021)]{LiYoudin2021}
Li, R., Youdin, A.~N., 2021.
Thresholds for particle clumping by the streaming instability.
The Astrophysical Journal 919, 107.

\bibitem[Lunine and Stevenson(1982)]{LunineStevenson1982}
Lunine, J.~I., Stevenson, D.~J., 1982.
Formation of the Galilean satellites in a gaseous nebula.
Icarus 52, 14--39.

\bibitem[Martin and Lubow(2011)]{MartinLubow2011}
Martin, R.~G., Lubow, S.~H., 2011.
Tidal truncation of circumplanetary discs.
Monthly Notices of the Royal Astronomical Society 413, 1447--1461.

\bibitem[Mosqueira and Estrada(2003)]{MosqueiraEstrada2003}
Mosqueira, I., Estrada, P.~R., 2003.
Formation of the regular satellites of giant planets in an extended
gaseous nebula I: subnebula model and accretion of satellites.
Icarus 163, 198--231.

\bibitem[Musiolik and Wurm(2019)]{MusiolikWurm2019}
Musiolik, G., Wurm, G., 2019.
Contacts of water ice in protoplanetary disks: Laboratory experiments.
The Astrophysical Journal 873, 58.

\bibitem[Nakagawa et al.(1986)]{NSH1986}
Nakagawa, Y., Sekiya, M., Hayashi, C., 1986.
Settling and growth of dust particles in a laminar phase of a low-mass
solar nebula.
Icarus 67, 375--390.

\bibitem[Nelson et al.(2013)]{NelsonGresselUmurhan2013}
Nelson, R.~P., Gressel, O., Umurhan, O.~M., 2013.
Linear and non-linear evolution of the vertical shear instability in
accretion discs.
Monthly Notices of the Royal Astronomical Society 435, 2610--2632.

\bibitem[Ormel and Cuzzi(2007)]{OrmelCuzzi2007}
Ormel, C.~W., Cuzzi, J.~N., 2007.
Closed-form expressions for particle relative velocities induced by
turbulence.
Astronomy \& Astrophysics 466, 413--420.

\bibitem[Pringle(1991)]{Pringle1991}
Pringle, J.~E., 1991.
The properties of external accretion discs.
Monthly Notices of the Royal Astronomical Society 248, 754--759.

\bibitem[Ronnet et al.(2017)]{Ronnet2017}
Ronnet, T., Mousis, O., Vernazza, P., 2017.
Pebble accretion at the origin of water in Europa.
The Astrophysical Journal 845, 92.

\bibitem[Schoonenberg and Ormel(2017)]{SchoonenbergOrmel2017}
Schoonenberg, D., Ormel, C.~W., 2017.
Planetesimal formation near the snowline: in or out?
Astronomy \& Astrophysics 602, A21.

\bibitem[Shakura and Sunyaev(1973)]{ShakuraSunyaev1973}
Shakura, N.~I., Sunyaev, R.~A., 1973.
Black holes in binary systems. {O}bservational appearance.
Astronomy \& Astrophysics 24, 337--355.

\bibitem[Shibaike et al.(2017)]{Shibaike2017}
Shibaike, Y., Okuzumi, S., Sasaki, T., Ida, S., 2017.
Satellitesimal formation via collisional dust growth in steady
circumplanetary disks.
The Astrophysical Journal 846, 81.

\bibitem[Shibaike et al.(2019)]{Shibaike2019}
Shibaike, Y., Ormel, C.~W., Ida, S., Okuzumi, S., Sasaki, T., 2019.
The Galilean satellites formed slowly from pebbles.
The Astrophysical Journal 885, 79.

\bibitem[Stevenson and Lunine(1988)]{StevensonLunine1988}
Stevenson, D.~J., Lunine, J.~I., 1988.
Rapid formation of Jupiter by diffusive redistribution of water vapor
in the solar nebula.
Icarus 75, 146--155.

\bibitem[Stoll and Kley(2014)]{StollKley2014}
Stoll, M.~H.~R., Kley, W., 2014.
Vertical shear instability in accretion disc models with radiation
transport.
Astronomy \& Astrophysics 572, A77.

\bibitem[Szul\'agyi et al.(2014)]{Szulagyi2014}
Szul\'agyi, J., Morbidelli, A., Crida, A., Masset, F., 2014.
Accretion of Jupiter-mass planets in the limit of vanishing
viscosity.
The Astrophysical Journal 782, 65.

\bibitem[Takeuchi and Lin(2002)]{TakeuchiLin2002}
Takeuchi, T., Lin, D.~N.~C., 2002.
Radial flow of dust particles in accretion disks.
The Astrophysical Journal 581, 1344--1355.

\bibitem[Tanigawa et al.(2012)]{Tanigawa2012}
Tanigawa, T., Ohtsuki, K., Machida, M.~N., 2012.
Distribution of accreting gas and angular momentum onto
circumplanetary disks.
The Astrophysical Journal 747, 47.

\bibitem[Teague et al.(2019)]{Teague2019}
Teague, R., Bae, J., Bergin, E.~A., 2019.
Meridional flows in the disk around a young star.
Nature 574, 378--381.

\bibitem[Tominaga and Tanaka(2025)]{TominagaTanaka2025}
Tominaga, R.~T., Tanaka, H., 2025.
Dust coagulation assisted by streaming instability in
protoplanetary disks.
The Astrophysical Journal 983, 15.

\bibitem[Turner et al.(2014)]{TurnerLeeSano2014}
Turner, N.~J., Lee, M.~H., Sano, T., 2014.
Magnetic coupling in the disks around young gas giant planets.
The Astrophysical Journal 783, 14.

\bibitem[Ueda et al.(2024)]{Ueda2024}
Ueda, T., Tazaki, R., Okuzumi, S., Flock, M., Sudarshan, P., 2024.
Support for fragile porous dust in a gravitationally self-regulated
disk around IM Lup.
Nature Astronomy 8, 1148--1158.

\bibitem[Wada et al.(2009)]{Wada2009}
Wada, K., Tanaka, H., Suyama, T., Kimura, H., Yamamoto, T., 2009.
Collisional growth conditions for dust aggregates.
The Astrophysical Journal 702, 1490--1501.

\bibitem[Weidenschilling(1977)]{Weidenschilling1977}
Weidenschilling, S.~J., 1977.
Aerodynamics of solid bodies in the solar nebula.
Monthly Notices of the Royal Astronomical Society 180, 57--70.

\bibitem[Xu and Bai(2022)]{XuBai2022}
Xu, Z., Bai, X.-N., 2022.
Dust settling and clumping in MRI-turbulent outer protoplanetary
disks.
The Astrophysical Journal 924, 3.

\bibitem[Yang et al.(2018)]{Yang2018}
Yang, C.-C., Mac Low, M.-M., Johansen, A., 2018.
Diffusion and concentration of solids in the dead zone of a
protoplanetary disk.
The Astrophysical Journal 868, 27.

\bibitem[Yap and Batygin(2025)]{YapBatygin2026}
Yap, T.~E., Batygin, K., 2025.
Callisto's nonresonant orbit as an outcome of circum-Jovian disk
substructure.
The Astrophysical Journal 995, 218.

\bibitem[Yap and Stevenson(2026)]{YapStevenson2026}
Yap, T.~E., Stevenson, D.~J., 2026.
Formation of water-rich giant planet satellites at decretion disk
ice lines.
The Planetary Science Journal 7, 44.

\end{thebibliography}
\end{document}